\documentclass[12pt]{iopart}
\usepackage{graphicx}

\begin{document}
\title[Anisotropy of DM annihilation]{Anisotropy of dark matter
annihilation with respect to the Galactic plane}

\author{V S Berezinsky$^1$, V I Dokuchaev$^2$ and
Yu N Eroshenko$^2$}

\address{$^1$ INFN, Laboratori Nazionali del Gran Sasso, I-67010
  Assergi (AQ), Italy}
\address{$^2$ Institute for Nuclear Research of the Russian Academy of
 Sciences, Moscow, Russia}
\ead{dokuchaev@inr.npd.ac.ru}

\begin{abstract}
We describe the anisotropy of dark matter clump distribution
caused by tidal destruction of clumps in the Galactic disk. A
tidal destruction of clumps with orbit planes near the disk plane
occurs more efficiently as compared with destruction of clumps at
near-polar orbits. A corresponding annihilation of dark matter
particles in small-scale clumps produces the anisotropic gamma-ray
signal with respect to the Galactic disk. This anisotropy is
rather small, $\sim9$\%, and superimposed on that due to
off-centering position of the Sun in the Galaxy. The anisotropy of
annihilation signal with respect to the Galactic disk provides a
possibility to discriminate dark matter annihilation from the
diffuse gamma-ray backgrounds of other origin.
\end{abstract}

\submitto{JCAP}

\maketitle

\section{Introduction}

A primordial power-law spectrum of density fluctuations in the
Dark Matter (DM) ranges from the largest scales above the scales
of superclusters of galaxies to the smallest sub-stellar scales
according to prediction of inflation models. This permits to
predict the properties of smallest DM structures from the known
CMB fluctuations at large scales. Substructures of DM in the
galactic haloes with a rather large mass, $\ge10^7M_{\odot}$, were
extensively discussed in early works, see for example
\cite{silk93}. The nonlinear dynamics and mechanism of
hierarchical clustering of these large DM clumps were analyzed in
both analytical calculations \cite{ufn1,ufn2,ufn3} and numerical
simulations \cite{nfw,ghigna98,moore99}. At sub-stellar
mass-scales of DM fluctuations, a principal new phenomenon arise
--- the cutoff of mass spectrum due to collisional and
collisionless (free streaming) damping processes of DM particles
in the forming clumps. The resulting smallest mass of DM clumps is
determined by the properties of DM particles, in particular, by
their elastic scattering. For detailed calculations of this cutoff
see e.~g. \cite{GreHofSch05} and references therein. Additionally
the cutoff of mass spectrum is influenced by the acoustic
absorption \cite{weinberg} at the time of kinetic decoupling of DM
particles \cite{bino} and also by the horizon-scale perturbation
modes \cite{LoeZal05}. In \cite{Bert06} the kinetic equations for
DM phase space density were solved in the case of perturbed
cosmological background by taking into account the acoustic
absorption, horizon-scale modes and gravitational perturbations. A
corresponding value of the smallest clump mass for neutralino DM
is of the order of the Moon or Earth mass.

The formation of small-scale DM clumps with a mass larger than the
Earth mass, $M_{\rm min}\sim10^{-6}M_{\odot}$, have been explored
in numerical simulations \cite{DieMooSta05,DieKuhMad06}. A
resulting differential number density of small-scale clumps,
$n(M)\,dM \propto dM/M^2$, turns out very close to that obtained
in the numerical simulations of large-scale clumps with mass
$M\geq10^6M_{\odot}$. The other important result obtained in
numerical simulations \cite{DieMooSta05} is determination of the
internal density profile in the isolated clump of minimal mass.
The resulting density profile is approximately a power-law,
$\propto r^{-\beta}$, with $\beta=1.5-2.0$, which is in a good
agreement with theoretically predicted value $\beta=1.7-1.8$
according to \cite{ufn1}.

The number density of small-scale DM clumps existing nowadays in
the Galactic halo is determined by their tidal destruction during
hierarchical structure formation \cite{bde03} and also by tidal
interactions with stars in the Galaxy
\cite{DieMooSta05,ZTSH,ZTSH0508,MDSQ,bde06,Green06,AngZha06}.
Annihilation of DM particles in small-scale clumps
\cite{DieKuhMad06,BBM97,ABO,olinto,silk02,berg98,berg2001,berg2002,Oda05}
enhances the total DM annihilation signals in our Galaxy and thus
boosts a chance for indirect detection of DM.

The usual assumption in calculations of DM annihilation is a
spherical symmetry of the Galactic halo. In this case an
anisotropy of annihilation gamma-radiation is only due to
off-center position of the Sun in the Galaxy. Nevertheless, a
principal significance of the halo nonsphericity for the observed
annihilation signal was demonstrated in \cite{CalMoo00}. According
to observations, the axes of the Galactic halo ellipsoid differ
most probably no more than $10-20$\%, but even a much more larger
difference of axes, up to a factor 2, can not be excluded
\cite{OllMer00,OllMer01}. This leads to more than an order of
magnitude uncertainty in the predicted annihilation flux from the
Galactic anti-center direction \cite{CalMoo00}. It must be noted
also the ``intrinsic'' annihilation anisotropy caused by the
small-scale DM clustering itself. A corresponding angular power
spectrum of annihilation signal at small scales is connected with
a power spectrum of DM clumping \cite{AndKom06}. In principle, the
DM clumps may be seen as point sources at the gamma-sky
\cite{CalMoo00}. Another minor source of annihilation anisotropy
is a dipole anisotropy due to proper motion of the Sun in the
Galaxy \cite{HooSer07}.

In \cite{ZTSH0508} the anisotropy with respect to the Galactic
disk was discussed basing on the numerical calculations of the
destruction of DM clumps by stars in the disk and taking into
account the influence of gravitational potential of the disk on
the clump orbits. It was also shown \cite{bde03,bde06} that (i)
small-scale DM clumps dominate in the generation of annihilation
signal and (ii) the Galactic stellar disk provides the main
contribution to the tidal destruction of clumps at $r>3$~kpc,
i.~e. outside the central bulge region. A process of clump
destruction in the halo is anisotropic in general (e.~g. it
depends on the inclination of clump orbit with respect to the disk
plane). Respectively, the DM annihilation in the halo is also
anisotropic. In this work we estimate the value of this
anisotropy. It must be stressed that with a present state of art
it is impossible to separate this source of anisotropy from that
produced by the halo nonsphericity. More detailed investigation is
required to constrain the shape of the halo and to search the
distinctive features of annihilation anisotropy due to
non-spherical halo clumpinesss.  The detectors at the GLAST
satellite will be sensitive to anisotropy up to $0.1$\% level
\cite{HooSer07}. This will provide a hope to discriminate the
anisotropic DM annihilation signal from the diffuse gamma-ray
backgrounds.

\section{Destruction of clumps by disk}
\label{disksec}

Crossing the Galactic disk, a DM clump can be tidally destructed
by the collective gravitational field of stars in the disk. This
phenomenon is similar to the destruction of globular clusters by
the ``tidal shocking'' in the Galactic disk \cite{OstSpiChe}. The
corresponding energy gain per unit mass of a clump at one disk
crossing \cite{OstSpiChe} is
\begin{equation}
  \label{disksh1}
 \Delta \tilde E=\frac{2g_m^2(\Delta z)^2}{v_{z,c}^2},
\end{equation}
where $g_m$ is the maximum gravitational acceleration of the clump
moving through the disk, $\Delta z$ is a vertical (perpendicular
to the disk plane) distance of a DM particle from the clump
center, $v_{z,c}$ is a vertical component of velocity at disk
crossing. The dependence of $v_{z,c}$ on the inclination of orbit
relative to the disk plane is the origin of the discussed
anisotropy in the clump destruction, and, as a result, the origin
of the anisotropy in annihilation signal.

The surface mass of the Galactic disk \cite{MarSuch} can be
approximated as
\begin{equation}
  \label{diskmass}
\sigma_s(r)=\frac{M_{\rm d}}{2\pi r_0^2}\,e^{-r/r_0},
\end{equation}
with $M_{\rm d}=8\times10^{10}M_{\odot}$ and $r_0=4.5$~kpc, and
therefore
\begin{equation}
 g_m(r)=2\pi G\sigma_s(r).
 \label{diskacc}
\end{equation}
We use the power-law parametrization
\cite{ufn1,ufn2,ufn3,DieMooSta05} of the internal density of a
clump
\begin{equation}
 \rho_{\rm int}(r)=
 \frac{3-\beta}{3}\,\rho\left(\frac{r}{R}\right)^{-\beta},
 \label{rho}
\end{equation}
where $\rho$ and $R$ are the mean internal density and a radius of
clump, respectively, $\beta=1.8$ and $\rho_{\rm int}(r)=0$ at
$r>R$. The total (kinetic plus potential) internal energy of a
clump for density profile (\ref{rho}) is given by
\begin{equation}
 |E|=\frac{3-\beta}{2(5-2\beta)}\frac{GM^2}{R},
 \label{Etot}
\end{equation}
where $M$ is the mass of the clump. Integrating (\ref{disksh1})
over a clump volume and using the density  profile (\ref{rho}), one
obtains  an energy gain for the whole clump as
\begin{equation}
 \frac{\Delta E}{|E|}=
 \frac{(5-2\beta)}{\pi(5-\beta)}\frac{g_m^2}{G\rho v_{z,c}^2}.
 \label{tdisk}
\end{equation}
We will use the following criterium for a tidal destruction of
clump: a clump is destructed if a total energy gain $\sum\Delta
E_i$ after several disk crossings exceeds the initial internal
energy $|E|$ of a clump.

Let us consider now some particular orbit of a clump in the halo
with an ``inclination'' angle $\gamma$ between the normal vectors
of the disk plane and orbit plane. The orbit angular velocity at a
distance $r$ from the Galactic center is $d\phi/dt=J/(mr^2)$,
where $J$ is an orbital angular momentum of a clump. A vertical
velocity of a clump crossing the disk is
\begin{equation}
 v_{z,c}=\frac{J}{mr_{\rm c}}\sin\gamma,
 \label{vzc}
\end{equation}
where $r_{\rm c}$ is a distance of crossing point from the Galaxy
center. There are two crossing points (with different values of
$r_{\rm c}$) during the one orbital period. The momentum
approximation used here for calculations of the tidal heating is
violated at small inclination angles, $\gamma\ll1$. Nevertheless
the resultant anisotropy is a cumulative quantity. It results from
an integration over all clump orbits, and orbits with $\gamma\ll1$
provide only small input into the anisotropy value.

We use here the Navarro-Frenk-White (NFW) density profile for the
Galactic halo:
\begin{equation}
 \rho_{\rm H}(r)=\frac{\rho_0}{(r/L)(1+r/L)^2},
 \label{nfwhalo}
\end{equation}
where $\rho_0$ is normalized in such a way that at position of the
Sun $\rho_{\rm H}(r_\odot)=0.3$~GeV/cm$^3$ and $L=45$~kpc.

Let us introduce the following set of dimensionless variables:
\begin{equation}
x=\frac{r}{L}, \quad \tilde\rho(x)=\frac{\rho_{\rm H}(r)}{\rho_0},
\quad y=\frac{J^2}{8\pi G\rho_0L^4M^2},
\end{equation}
\begin{equation}
\varepsilon=\frac{E_{\rm orb}/M-\Phi_0}{4\pi G\rho_0L^2 }, \quad
\psi=\frac{\Phi-\Phi_0}{4\pi G\rho_0L^2},
\end{equation}
where $\Phi_0=-4\pi G\rho_0L^2$, $E_{\rm orb}$ is an total orbital
energy (kinetic and potential) of a clump. In these dimensionless
variables the density profile of the halo (\ref{nfwhalo}) has a
form
\begin{equation}
\tilde\rho(x)=\frac{1}{x(1+x)^2}.
 \label{nfwhalox}
\end{equation}
Consider the Galactic halo model with an isotropic velocity
distribution, which is appropriate for the halo formed by the
hierarchial clustering of sub-haloes. In this model the
distribution function depends only on energy. The gravitational
potential $\psi(x)$, which corresponds to the density profile
(\ref{nfwhalox}) is given by
\begin{equation}
 \psi(x)=1-\frac{\log(1+x)}{x}.
 \label{pot}
\end{equation}
The equation for the turning points of an orbit, $\dot r^2=0$, for
the potential (\ref{pot}) can be written as
\begin{equation}
1-\frac{\log(1+x)}{x}+\frac{y}{x^2}=\varepsilon. \label{turn}
\end{equation}
From (\ref{turn}) one can find numerically the minimum $x_{\rm
min}$ and maximum $x_{\rm max}$ distance between a clump and the
Galactic center as function of $\varepsilon$ and $y$. Now denoting
$p=\cos\theta$, where $\theta$ is an angle between the
radius-vector $\vec r$ and the particle velocity $\vec v$, we have
\begin{equation}
 y=(1-p^2)\,x^2\left(\varepsilon-\psi(x)\right)
 \label{yexp}.
\end{equation}
According to our assumption, the vectors $\vec v/v$ are
distributed isotropically at each point $x$, and therefore $p$ has
a uniform distribution in the interval $[0,1]$. The relation
between $\tilde\rho(x)$ and the distribution function
$F(\varepsilon)$ is given, according to \cite{Edd16}, by
\begin{equation}
\tilde\rho(x)=4\pi\sqrt{2}\int\limits_{\psi(x)}^{1}d\varepsilon
\left[\varepsilon-\psi(x)\right]^{1/2}F(\varepsilon).
 \label{rhofd}
\end{equation}
The function $F(\varepsilon)$ for a halo profile (\ref{nfwhalox})
can be fitted \cite{Wid00} with a good accuracy as
\begin{equation}
F(\varepsilon)=F_1(1-\varepsilon)^{3/2}\varepsilon^{-5/2}
\left[-\frac{\ln(1-\varepsilon)}{\varepsilon}\right]^qe^P,
 \label{fffrhofd}
\end{equation}
where $F_1=9.1968\times10^{-2}$,
$P=\sum\limits_{i}p_i(1-\varepsilon)^i$,
$(q{,}p_1{,}p_2{,}p_3{,}p_4)=(-2.7419{,} 0.3620{,} -0.5639{,}
-0.0859{,} -0.4912)$.

The time of motion from $x_{\rm min}$ to $x_{\rm max}$ and back is
\begin{equation}
 T_{\rm c}(x,\varepsilon,p)=\frac{1}{\sqrt{2\pi G\rho_0}}
 \int\limits_{x_{\rm min}}^{x_{\rm max}}
 \frac{ds}{\sqrt{\varepsilon-\psi(s)-y/s^2}},
 \label{pcint}
\end{equation}
where $y$ depends on $x$, $\varepsilon$ and $p$ according to
(\ref{yexp}). The trajectory (the orbit) of a clump in the
potential (\ref{pot}) is not closed. A precession angle during the
time $T_{\rm c}/2$ is
\begin{equation}
 \tilde\phi=y^{1/2}\int\limits_{x_{\rm min}}^{x_{\rm max}}
 \frac{ds}{s^2\sqrt{\varepsilon-\psi(s)-y/s^2}}-\pi,
 \label{prec}
\end{equation}
and $\tilde\phi<0$. Therefore the orbital period is greater then
$T_{\rm c}$ and equals to
\begin{equation}
 T_{\rm t}=T_{\rm c}\left(1+\tilde\phi/\pi\right)^{-1}.
 \label{pcintt}
\end{equation}
During a life-time of the Galaxy, $t_{\rm G}\simeq10^{10}$~yrs,
the relative total energy gain of a clump due to the tidal heating
in multiple crossings of the disk is
\begin{equation}
 \frac{\Delta E}{|E|}=\frac{1}{|E|}
 \sum\limits_{i=1}^{N}(\Delta E^{i}_1+\Delta E^{i}_2),
\end{equation}
where $\Delta E^{i}_1$ and $\Delta E^{i}_2$ are given by
(\ref{tdisk}) for the two disk crossings during an orbital period
$T_{\rm t}$, and $N\simeq t_{\rm G}/T_{\rm t}$. According to
(\ref{vzc}) the $z$-velocity component $v_{z,c}\propto x_{\rm
c}^{-1}$, where $x_{\rm c}=r_{\rm c}/L$. Therefore one has to
calculate the sum $\sum g_m^2(x_{\rm c})x_{\rm c}^2$ with
summation over all subsequent crossing points (odd and even) of a
clump orbit with the Galactic disk. An important simplification in
calculations follows from the fact that precession velocity is
constant. For this reason the points of successive odd crossings
are separated by the same angles $\tilde\phi$. The same is also
true for the successive even crossings. Using this simplification
one can do the following transformation:
\begin{equation}
 \sum\limits_{i=1}^{N}g_m(x)^2x^2\simeq
\frac{1}{|\tilde\phi|}\int \!g_m(x)^2x^2d\phi
 \simeq\frac{1}{|\tilde\phi|}\int\limits_{x_{\rm min}}^{x_{\rm max}}
 \!g_m(x)^2x^2\frac{d\phi}{dx}\,dx\frac{2t_{\rm G}}{T_{\rm t}},
 \nonumber
\end{equation}
where
\begin{equation}
 \frac{d\phi}{dx}=
 \frac{y^{1/2}}{x^2\sqrt{\varepsilon-\psi(x)-y/x^2}}
\end{equation}
is an equation of clump trajectory in the halo. Using the above
formulae one finds
\begin{equation}
 \Delta E/|E|=\frac{A}{\sin^2\gamma},
 \label{a1}
\end{equation}
where
\begin{equation}
A(x,\varepsilon,y)=\frac{(5-2\beta)\,t_{\rm G}}{2\pi^2(5-\beta)
G^2\rho\rho_0L^2y^{1/2}|\tilde\phi|T_{\rm t}}\int\limits_{x_{\rm
min}}^{x_{\rm max}}
 \frac{g_m^2(s)\,ds}{\sqrt{\varepsilon-\psi(s)-y/s^2}}.
 \nonumber
 \label{a2}
\end{equation}

\section{Anisotropy of clump distribution}

A tidal heating and final destruction of clumps by the
gravitational field of the Galactic disk depends on the
inclination angle $\gamma$ of a clump orbit to the disk according
to (\ref{disksh1}) and (\ref{a1}). This is a cause of the
anisotropic clump number density decreasing during the lifetime of
the Galaxy. A tidally induced anisotropy of clump distribution can
be taken into account by inserting into the integral (\ref{rhofd})
an additional factor $e^{-\Delta E/|E|}=e^{-t_{\rm G}/t_{\rm d}}$,
where $t_{\rm d}$ is an effective time of clump destruction. Due
to an ``anisotropy'' factor $1/\sin^2\gamma$ in (\ref{a1}) the
``tidal shocking'' by the Galactic disk is most effective for
clumps with orbit planes which are near coplanar to the disk plane
(that is with an inclination $\gamma\ll1$). Meanwhile for any
intersection points there are many others orbits which pass
through the disk plane with $\gamma\sim1$. For this reason the
resulting anisotropy in clump distribution is rather small.

\begin{figure}[t]
\includegraphics[width=0.95\textwidth]{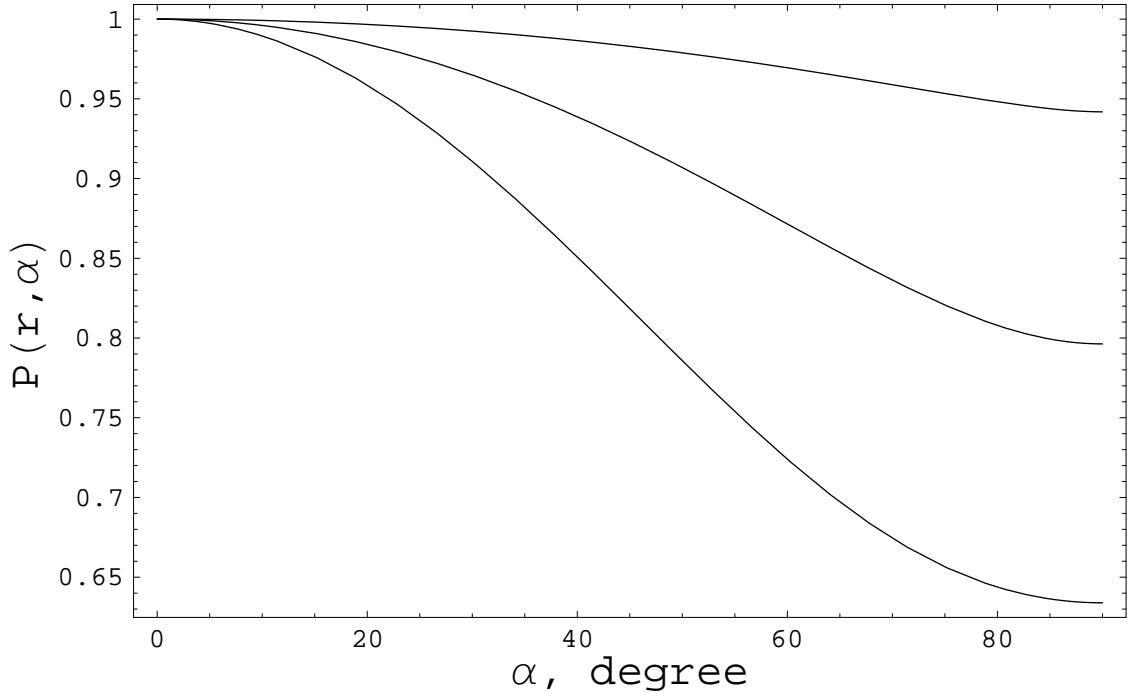}
\caption{The normalized fractions of DM clumps in the halo
$P(r,\alpha)$ from (\ref{sp1}) which survive the tidal destruction
by the stellar disk as a function of angle $\alpha$ between a
radius-vector $\vec r$ and the disk polar axis. The plots are
shown for radial distances from the Galactic center $r=3$, $8.5$
and $20$~kpc (from the bottom to the top). The curves must be
multiplied by factors 0.04, 0.4 and 0.9 respectively to reproduce
the actual values.}
 \label{ani1f}
\end{figure}

Let us consider a point in the halo with a radius vector $\vec r$
and an angle $\alpha$ with a polar axis of the disk. Only orbits
with $\pi/2-\alpha<\gamma<\pi/2$ go through this point. A survival
probability for clumps can be written now in the following form
\begin{equation}
 P(x,\alpha)\!=\!\frac{4\pi\sqrt{2}}{\tilde\rho(x)\sin\alpha}
 \int\limits_{0}^{1}dp\int\limits_{0}^{\sin\alpha}d\cos\gamma
 \int\limits_{\psi(x)}^{1}\!d\varepsilon\,
 [\varepsilon-\psi(x)]^{1/2}F(\varepsilon)\,e^{-\Delta E/|E|},
 \nonumber
 \label{sp1}
\end{equation}
where $\Delta E/|E|$ is defined by (\ref{a1}). This expression is
derived from (\ref{rhofd}) by inserting the distribution over
additional parameters $p$ and $\gamma$ (by taking in mind that we
use the isotropic halo model) and the exponential factor for
clumps destruction $e^{-\Delta E/|E|}$. The numerically calculated
triple integral from (\ref{sp1}) for survival probability
$P(r,\alpha)$ is shown in the Fig.~{\ref{ani1f}}.  The
annihilation anisotropy is artificially enhanced in the
Fig.~{\ref{ani1f}} for better visualization for three chosen
radial distances from the Galactic center, $r=3$, $8.5$ and
$20$~kpc respectively by using the different multiplication
factors.

\section{Annihilation anisotropy}

For the diffuse distribution of DM in the halo, the annihilation
signal (e.~g. gamma-ray or neutrino flux per unit solid angle) is
proportional to
\begin{equation}
I_{\rm H}=\int\limits_{0}^{r_{\rm max}(\zeta)}\rho_{\rm
H}^2(\xi)\,dx,
 \label{ihal1}
\end{equation}
where $x=r/L$ and integration over $r$ goes along the line of
sight, $\xi(\zeta,r) =
(r^2+r_{\odot}^2-2rr_{\odot}\cos\zeta)^{1/2}$ is the distance to
the Galactic center, $r_{\rm max}(\zeta) = (R_{\rm
H}^2-r_{\odot}^2\sin^2\zeta)^{1/2} + r_{\odot}\cos\zeta$  is the
distance to the external halo border, $\zeta$ is an angle between
the line of observation and the direction to the Galactic center,
$R_{\rm H}$ is a virial radius of the Galactic halo,
$r_{\odot}=8.5$~kpc is the distance between the Sun and Galactic
center. The corresponding signal from annihilations in DM clumps
is proportional to the quantity \cite{bde03}
\begin{equation}
 I_{\rm cl}=\mu S\rho\int\limits_{0}^{r_{\rm max}(\zeta)}
 \rho_{\rm H}(\xi)P(\xi,\alpha)P_{\rm sp}(\xi)\,dx,
 \label{ihal2}
\end{equation}
where $\mu\simeq0.05$ is a fraction of the DM mass in the form of
clumps, $P_{\rm sp}$ is a survival probability of clumps due to
their tidal destructions by stars in the halo and bulge from
\cite{bde06}. The function $S$ depends on the clump density
profile and core radius of clump \cite{bde03} and we use $S\simeq
14.5$. Here for simplicity we do not take into account the
distribution of DM clumps over their internal densities. As a
representative example we consider the Earth-mass clumps
$M=10^{-6}M_{\odot}$ originated from $2\sigma$ density peaks in
the case of power-law index of primordial spectrum of
perturbations $n_p=1$. The mean internal density of these clumps
is $\rho\simeq7\times10^{-23}$~g~cm$^{-3}$. The  values of $\mu$
and $S$, as well as the distribution of the clumps over various
parameters influence the annihilation signal but only weakly
affect the predicted anisotropy.
\begin{figure}[t]
\includegraphics[width=0.95\textwidth]{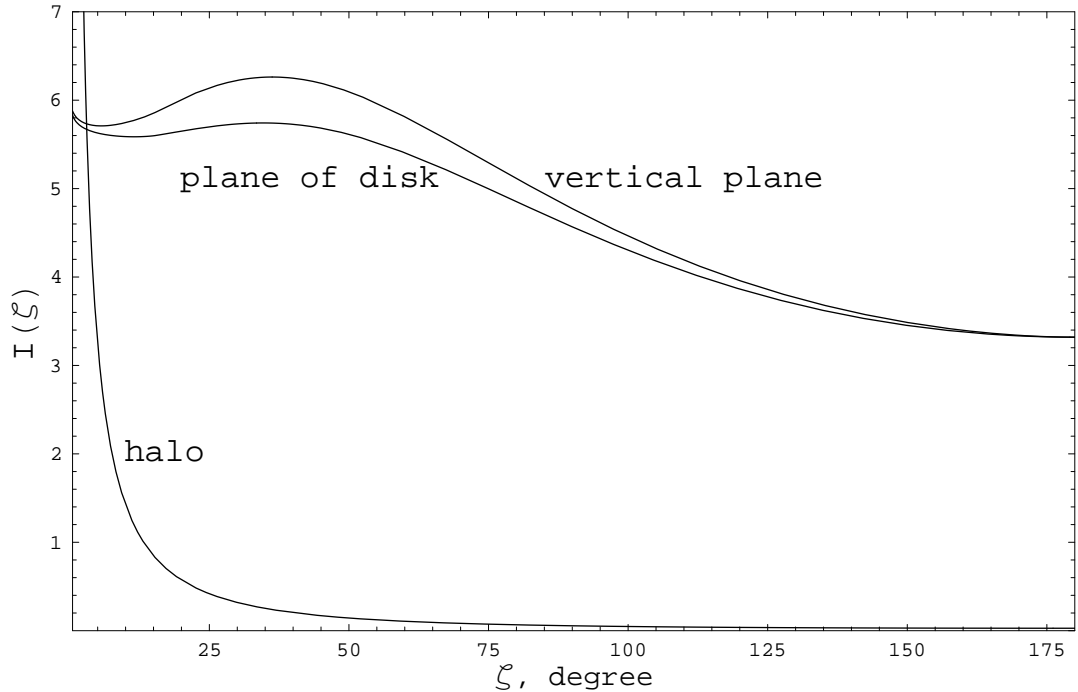}
\caption{The annihilation signal (\ref{ihal2}) in the Galactic
disk plane and in vertical plane as a function of the angle
$\zeta$ between the line of observation and the direction to the
Galactic center. For comparison it is shown also the annihilation
signal from the Galactic halo without DM clumps (\ref{ihal1}). The
values of both integrals (\ref{ihal2}) and (\ref{ihal1}) are
multiplied by factor $10^{48}$. }
 \label{ani2f}
\end{figure}

In the Fig.~\ref{ani2f} the annihilation signal calculated
according to (\ref{ihal2}) is shown for the Galactic disk plane
and for the orthogonal vertical plane (passing through the
Galactic center) as function of angle $\zeta$ between the
observation direction and the direction to the Galactic center.
For comparison in the Fig.~\ref{ani2f} is also shown the signal
from the spherically symmetric Galactic halo without the DM clumps
(\ref{ihal1}). The later signal is the same in the in the Galactic
disk plane and in vertical plane and therefore can be principally
extracted from the observations.

The difference of the signals in two orthogonal planes at the same
$\zeta$ can be considered as an anisotropy  measure. Defined as
$\delta=(I_2-I_1)/I_1$, it has a maximum value $\delta \simeq
0.09$ at  $\zeta\simeq39^{\circ}$.

\section{Discussions}
 \label{discussion}

A total anisotropy of DM annihilation signal is determined in
general by the Sun off-centering in the Galaxy and by the halo
nonsphericity. An annihilation signal from the Galactic center
depends on the central density profile of the Galactic halo. In
the case with a density cusp \cite{BGZ}, the bright source in the
Galactic center is inevitable. Meanwhile, this cusp in the diffuse
DM may be destroyed by the stellar feedback \cite{Mash06}. The
small-scale DM clumps are completely destructed inside the
Galactic stellar bulge region. The ``gamma-rings'' are predicted
in other galaxies due to the absence of clumps in their centers
\cite{ZTSH0508}. The unknown nonsphericity of the halo is a main
source of anisotropy uncertainty. The value of anisotropy due to
nonsphericity of the halo may be several times larger than one
caused by the discussed in this paper effect of tidal destruction
of DM clumps by the disk.  More detailed analysis is required to
separate these two sources of anisotropy. A nonsphericity
(oblateness) of the halo due to the angular momentum can be easily
estimated. It is natural to assume that the DM halo and disk have
the same value of specific angular momentum (i.~e. an angular
momentum per unit mass). In this case the model of the Maclaurin
spheroid for the halo gives only $\sim0.5$\% difference for the
halo axes. Therefore, the nonsphericity of the halo due to the
angular momentum produces a negligible anisotropy.

In \cite{ZTSH0508} the anisotropy with respect to disk plane in
the Galaxy was pointed out. As it is seen from our calculations
(see Fig.~\ref{ani2f}) this anisotropy of annihilation signal from
the DM clumps with respect to the disk is rather small, $\sim9$\%,
but far exceeds the anticipated GLAST resolution, $\sim0.1$\%.
Therefore, the discussed anisotropy may be used in future detailed
gamma-ray observations for discrimination of the annihilation
signals from the DM clumps, diffuse DM in the Galactic halo and
diffuse gamma-ray backgrounds.

In conclusion we list some unsolved problems and unclear features
of DM clump physics. First of all, the detailed analytical theory
of small DM clump clustering is desirable. An effective index of
the density perturbation power spectrum $n\to -3$ at small-scales.
This means that a gravitational clustering of small-scale
structures proceeds very fast. As a result, the formation of
small-scale DM clumps and their capturing by the larger ones are
nearly simultaneous processes. The only approach developed to
track this clustering is an approximate theoretical model
\cite{bde03,bde06} and restricted numerical simulations
\cite{DieMooSta05}. During the past years the minimum clump mass
is widely discussed. A value of the minimum mass has a principal
significance for annihilation signal calculations. For this
reason, the detailed calculations of minimum clump are requested.
The central parts of DM clumps dominates in the generation of
annihilation signal. For this reason the crucial problem is a
value of the central density of DM clumps. To recover this value a
detailed theoretical models and/or high-resolution numerical
simulations are needed. The other problem is a survival of the
central core of small-scale DM clump in tidal interactions. It is
possible in principle that survived cores of DM clumps dominate in
the annihilation signal. In this paper we evaluated only one
particular part of the clump anisotropy in the halo. It would be
useful also to clarify the halo shape and the nonsphericity of the
distribution of clump orbits due to influence of the gravitational
potential of the Galactic disk.

\ack

We thank Michael Kachelriess for useful discussion. This work was
supported in part by the Russian Foundation for Basic Research
grants 06-02-16029 and 06-02-16342, and the Russian President
grants LSS 4407.2006.2 and LSS 5573.2006.2.


\section*{References}

\begin{thebibliography}{99}

\bibitem{silk93} Silk J and Stebbins A 1993 {\it
Astrophys. J.} {\bf 411} 439

\bibitem{ufn1} Gurevich A V and Zybin K P 1988 {\it Sov.
Phys.--JETP} {\bf 67} 1

\bibitem{ufn2} Gurevich A V and Zybin K P  1988 {\it Sov.
Phys.--JETP} {\bf 67} 1957

\bibitem{ufn3} Gurevich A V and Zybin K P 1995 {\it Sov.
Phys.--Usp.} {\bf 165} 723

\bibitem{nfw} Navarro J F, Frenk C S and White S D M 1996 {\it
Astrophys. J.} {\bf 462} 563

\bibitem{ghigna98} Ghigna S, Moore B, Governato F, Lake G, Quinn T
and Stadel J 2000  {\it Astrophys. J.} {\bf 544} 616

\bibitem{moore99} Moore B et al. 1999 {\it Astrophys.
J.} {\bf 524} L19

\bibitem{GreHofSch05} Green A M Hofmann S and Schwarz D J 2005
{\it JCAP} {\bf 0508} 003

\bibitem{weinberg} Weinberg S 1971 {\it Astrophys.
J.} 168 175

\bibitem{bino}Hofmann S, Schwarz D J and Stocker H 2001 \PR D {\bf
64} 083507

\bibitem{LoeZal05} Loeb A and Zaldarriaga M 2005 \PR D {\bf
71} 103520

\bibitem{Bert06}Bertschinger E 2006 \PR D {\bf 74} 063509

\bibitem{DieMooSta05} Diemand J, Moore B and Stadel J 2005 {\it
Nature} {\bf 433} 389

\bibitem{DieKuhMad06} Diemand J, Kuhlen M and
Madau P 2006 {\it Astrophys. J.} {\bf 649} 1; {\it ibid.} 2007
{\it Astrophys. J.} {\bf 657} 262

\bibitem{bde03} Berezinsky V, Dokuchaev V and Eroshenko Yu 2003
\PR  D {\bf 68} 103003

\bibitem{ZTSH} Zhao H S, Taylor J, Silk J and Hooper D 2005
Earth-mass dark halos are torn into dark mini-streams by stars
{\it Preprint} astro-ph/0502049

\bibitem{ZTSH0508} Zhao H S, Taylor J, Silk J and Hooper D 2007
{\it Astrophys. J.} {\bf 654} 697

\bibitem{MDSQ} Moore B, Diemand J, Stadel J and Quinn T 2005 On
the survival and disruption of Earth mass CDM micro-haloes {\it
Preprint} astro-ph/0502213

\bibitem{bde06} Berezinsky V, Dokuchaev V and Eroshenko Yu 2006
\PR  D {\bf 73} 063504

\bibitem{Green06} Green A M and Goodwin S P 2007
{\it Mon. Not. Roy. Astron. Soc.} {\bf 375} 1111

\bibitem{AngZha06} Angus G W and Zhao H S 2006 Analysis of
galactic tides and stars on CDM microhalos {\it Preprint}
astro-ph/0608580

\bibitem{BBM97} Berezinsky V, Bottino A and Mignola G 1997 \PL B
{\bf 391} 355

\bibitem{ABO} Aloisio R, Blasi P and Olinto A V 2004
{\it Astrophys. J.} {\bf 601} 47

\bibitem{olinto} Tasitsiomi A and Olinto A V 2002 \PR D {\bf 66}
083006

\bibitem{silk02} Taylor J E and Silk J 2003
{\it Mon. Not. Roy. Astron. Soc.} {\bf 339} 505

\bibitem{berg98} Bergstrom L, Edsjo J and Ulio P 1998 \PR D {\bf
58} 083507

\bibitem{berg2001} Bergstrom L, Edsjo J and Gunnarsson C 2001
\PR D {\bf 63} 083515

\bibitem{berg2002} Ullio P, Bergstrom L, Edsjo J and Lacey C 2002
\PR D {\bf 66} 123502

\bibitem{Oda05} Oda T, Totani T and Nagashima M 2005 {\it
Astrophys. J.} {\it 633} L65

\bibitem{CalMoo00} Calcaneo-Roldan and Moore B 2000
\PR  D {\bf 62} 123005

\bibitem{OllMer00} Olling R P and Merrifield M R 2000 Mon. Not.
Roy. Astron. Soc. {\bf 311} 361

\bibitem{OllMer01} Olling R P and Merrifield M R 2001 Mon. Not.
Roy. Astron. Soc. {\bf 326} 164

\bibitem{AndKom06}Ando S and Komatsu E 2006 \PR D
{\bf 73} 023521

\bibitem{HooSer07} Hooper D and Serpico P D 2007
Angular Signatures of Dark Matter in the Diffuse Gamma Ray
Spectrum {\it Preprint} astro-ph/0702328

\bibitem{OstSpiChe} Ostriker J P, Spitzer L Jr and
Chevalier R A 1972 {\it Astrophys. J. Lett.} {\bf 176} 51

\bibitem{MarSuch} Marochnik L S and Suchkov A A 1984 {\it Galaxy}
(Moscow: Nauka)

\bibitem{Edd16} Eddington A S 1916 {\it Mon. Not. Roy. Astron. Soc.}
{\bf 76} 572

\bibitem{Wid00} Widrow L M 2000 {\it Astrophys. J. Supp.} {\bf 131} 39

\bibitem{BGZ} Berezinsky V S, Gurevich A V and Zybin K P 1992
\PL B {\bf 294} 221

\bibitem{Mash06} Mashchenko S, Couchman H M P and Wadsley J 2005
{\it Nature} {\bf 442} 539

\end{thebibliography}
\end{document}